\documentclass[twocolumn,showpacs,preprintnumbers,amsmath,amssymb]{revtex4}

\usepackage{graphicx}% Include figure files
\usepackage{dcolumn}% Align table columns on decimal point
\usepackage{bm}% bold math

\begin{document}

\preprint{}

\title{Observation of Buried Phosphorus Dopants near Clean Si(100)-(2$\times$1) with Scanning Tunneling Microscopy}

\author{Geoffrey W. Brown}
 \email{geoffb@lanl.gov}
\author{Holger Grube}%
\author{Marilyn E. Hawley}

\affiliation{
Materials Science \& Technology Division, Los Alamos National Laboratory, Los Alamos, NM 87545}

\date{May 24, 2004}% It is always \today, today,

\begin{abstract}
We have used scanning tunneling microscopy to identify individual phosphorus dopant atoms near the clean silicon (100)-(2$\times$1) reconstructed surface. The charge-induced band bending signature associated with the dopants shows up as an enhancement in both filled and empty states and is consistent with the appearance of \textit{n}-type dopants on compound semiconductor surfaces and passivated Si(100)-(2$\times$1).  We observe dopants at different depths and see a strong dependence of the signature on the magnitude of the sample voltage.  Our results suggest that, on this clean surface, the antibonding surface state band acts as an extension of the bulk conduction band into the gap.  The positively charged dimer vacancies that have been observed previously appear as depressions in the filled states, as opposed to enhancements, because they disrupt these surface bands.
\end{abstract}


\maketitle

\section{\label{sec:level1}Introduction}

Information about the nature of defects and dopants at the (100) surface of silicon is important because it impacts many areas including surface quantum electronic devices \cite{TuckerShen00, Ladd02, OBrien01}, shrinking conventional electronics, and processing steps for device fabrication\cite{Ditchfield00}.  In any of these applications, knowledge of the distribution and electronic nature of the defects and dopants would be useful in modeling the environment of, or the process that leads to, the final device.  As devices shrink, the local electronic environment is especially important since, at very small length scales, a single charged defect can induce large electric fields and disrupt device operation through effects such as random telegraph noise.  Recently we have begun to study these issues by showing that it is possible to observe charged defects on the clean (100)-(2$\times$1) surfaces of both \textit{n}-type and \textit{p}-type silicon using scanning tunneling microscopy (STM)\cite{GWBrown02,GWBrown03}.  Work by others has also shown charged defects and dopants on this surface after it has been hydrogen terminated\cite{Lyding01,Lyding02}.  The hydrogen termination moves the electronic surface states out of the bandgap.  In this paper we report the observation of phosphorus dopants near clean \textit{n}-type Si(100)-(2$\times$1) surfaces using STM.  This is an important result since hydrogen termination may not always be compatible with processing steps in a particular application.   

STM has been used extensively in the past on compound semiconductors to identify charged vacancies and dopants near the surface.  The mechanism is well understood and the data can be used to determine whether the features are positively or negatively charged and in some cases how far below the surface they are located \cite{EbertLong}.  In the STM images the features are visible as long range enhancements or depressions that are caused by charge-induced band bending that changes the local density states (LDOS) between the tip and sample Fermi levels.
Our previous work on charged defects on silicon showed results similar to those observed on compound semiconductor surfaces even though the antibonding states of the Si(100)-(2$\times$1) surface protrude into the bulk bandgap.  The images showed positively charged defects on both \textit{n}- and \textit{p}-type material, but with an absence of the filled-state signature on \textit{n}-type samples. This was attributed to additional tip-induced band bending from the tungsten tips ($\sim$ 4 eV workfunction) and low conductivity samples used in those experiments.  Downward band bending (from positive charge) was prevented during filled-state imaging because of the tip-induced band bending, always present under negative sample bias, that pinned the conduction band minimum against the Fermi level.  Consistent with this interpretation, the filled state depressions were observed on \textit{p}-type samples, in which the Fermi level is farther from the conduction band edge. 

While it was straightforward to observe charged defects in our low bias images of those samples, it was much more difficult to observe dopants on either material and only anecdotal evidence was obtained.  The difficulty was primarily due to a combination of the low dopant density expected for those samples ( $\sim$ 1 in every 1000 nm x 1000 nm area) and the lack of a signature in the filled states.  An obvious way to increase the density is to use higher conductivity samples but this will cause Fermi level pinning at even smaller gap voltages with tungsten tips since the sample's Fermi level will be closer to the conduction band edge.  However, tips of higher workfunction material will move the tip's chemical potential closer to the sample's valence band edge, potentially eliminating the pinning during low-bias filled state imaging.  
In this work we show that platinum/iridium tips ($\sim$ 5.5 eV workfunction) and more conductive samples allow us to observe charged phosphorus dopants below the surface of the silicon.  The results are again consistent with those from compound semiconductor and hydrogen passivated Si(100)-(2$\times$1) surfaces.  The dopant-induced features appear as long range enhancements in both filled- and empty-state images and differing strength signatures are seen, implying that they are buried at different depths.

\section{Experiment, Results and Discussion}

Our experiments were carried out on clean (100)-(2$\times$1) surfaces of prime grade \textit{n}-type silicon wafers doped with phosphorus at a density of $\sim$ 5 x 10$^{17}$ / cm$^{3}$.  Constant current STM imaging was carried out in an ultra-high vacuum chamber with a base pressure of ~ 3 x 10$^{-11}$ Torr.  Sample biases were kept between $\pm$ 1.2 volts for charge imaging. The silicon samples were cut and mounted using nickel-free tools and then transferred to ultra-high vacuum without any chemical cleaning in our laboratory.  After degassing below 600$^{\circ}$ C for several hours, the samples were flashed at $\sim$1250$^{\circ}$ C for 1 minute, cooled rapidly to $\sim$ 950$^{\circ}$C, and then cooled to room temperature over a period of 5 minutes. Electrochemically etched platinum/iridium alloy tips (90/10) were used for imaging.
	
Figure 1 shows images of the features that we attribute to subsurface dopants.  Figures 1a and 1b are simultaneously acquired filled- and empty-state images, respectively, of a feature that appears as a long range enhancement in both biases, but without a corresponding surface defect.  For comparison, a positively charged multi-dimer vacancy defect appears in the lower right section of the image.  As expected for these PtIr tips, the filled state depression associated with the positively charged surface vacancy is visible.  The extent of the dopant signature is similar to that of the charged defects but the amplitude is somewhat weaker in the empty states.  

\begin{figure}[htb]
\includegraphics[width=3in]{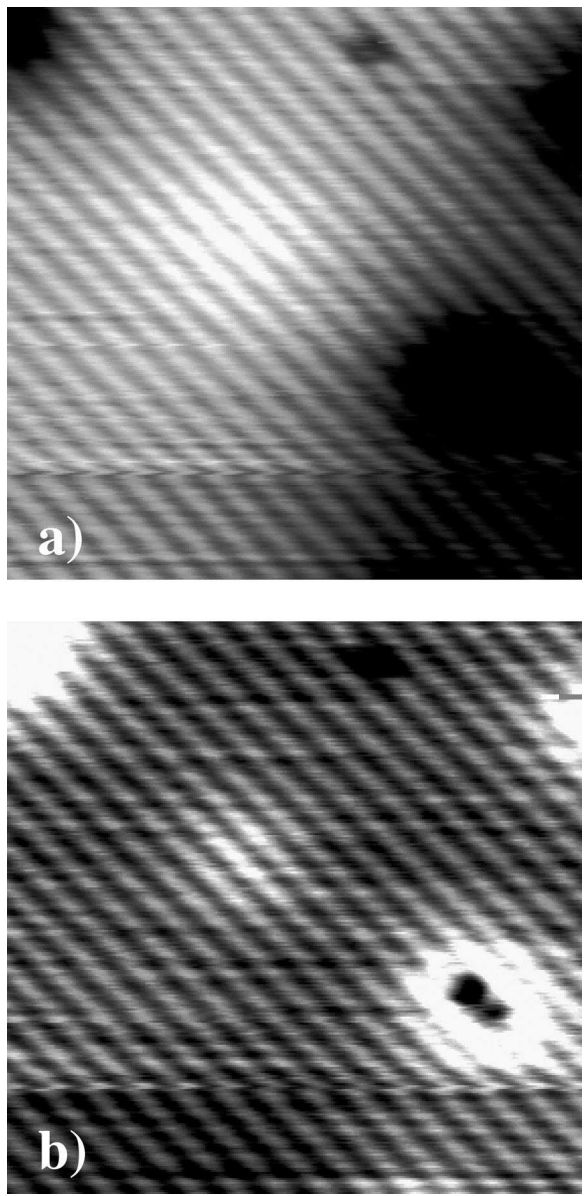}
\caption{
a) Filled- and b) empty-state images of the same area on a Si(100)-(2$\times$1) surface.  The images show a feature in the center which appears as a long range enhancement in both biases but which does not have an associated surface structure.  The filled states image was acquired at -0.7 V sample bias and the empty states at +1.0 V. The tunnel current for both was 130 pA and the image area is 100 $\AA$ $\times$ 100 $\AA$.
}
\label{fig:Fig1}
\end{figure}

Figures 2a and 2b show another one of these features at low (-0.6 V) and high (-1.5 V) sample bias, respectively.  The images were taken sequentially in the same area, as seen from the other defects.  In Figure 2b we clearly see the absence of a surface defect, indicating that the signature in Figure 2a arises from a subsurface feature.  At low bias, the band edges contribute strongly to the tunnel current and the effect of band bending is visible.  At high sample bias the band edge states contribute only very weakly to the tunnel current.

\begin{figure}[htb]
\includegraphics[width=3in]{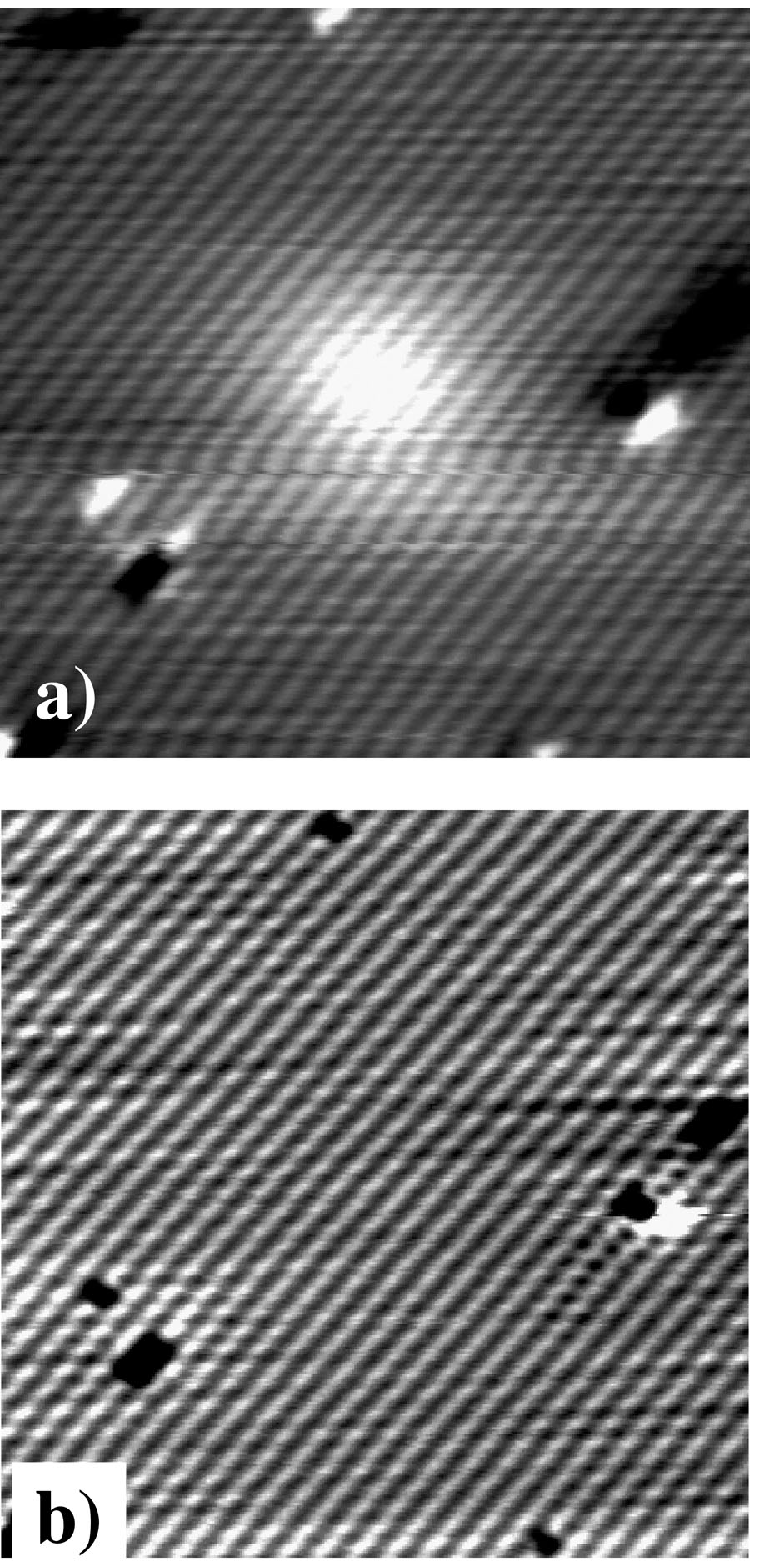}
\caption{Filled-states image of a subsurface dopant acquired at a) -0.6 V and b) -1.5 V sample bias.  The dopant signature has a strong bias dependence and is not visible at -1.5 V.  There is also no topographic surface structure (i.e. vacancy or adsorbate) associated with the feature. Image areas are both 220 $\AA$ $\times$ 220 $\AA$ and the tunnel currents were 110 pA.}
\label{fig:Fig2}
\end{figure}

Finally, in Figure 3 we see two of the dopant features in a single filled state image.  The upper feature has weaker amplitude than the lower one, implying that the upper dopant is either farther below the surface or has a smaller charge.  

\begin{figure}[htb]
\includegraphics[width=3in]{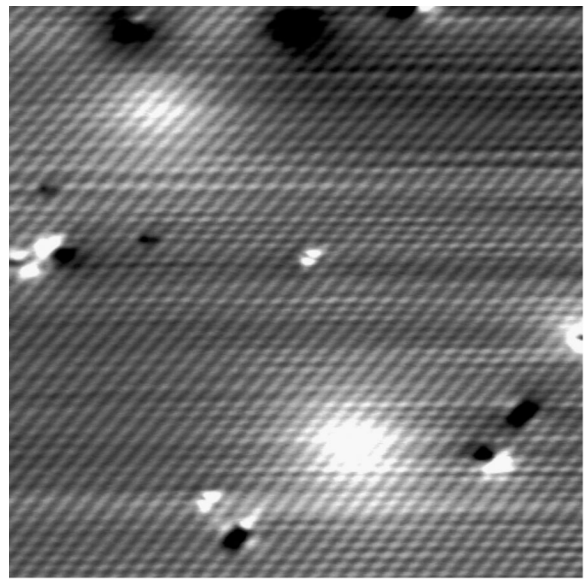}
\caption{Filled-state images of a pair of dopants on Si(100)-(2$\times$1).  The upper feature is weaker than the lower feature implying that it is either farther below the surface or has a smaller charge.  The tunnel current in both cases was 110 pA, the sample bias was -0.7 V and the image size is 320 $\AA$ $\times$ 320 $\AA$.}
\label{fig:Fig3}
\end{figure}

The interpretation of these features as phosphorus-dopant-induced is based on their density, the lack of other possible sources, and their appearance compared to dopants on \textit{n}-type compound semiconductor surfaces.  The density of the features in our images is about 1 for every 250 nm by 250 nm area or roughly half of what we might expect for dopants in the first three layers of silicon doped at ~ 5 x 10$^{17}$ / cm$^{3}$.  This is reasonable given the small number of features observed, the thermal processing steps involved in sample preparation, and the uncertainty in the doping level itself.  We choose three layers as a cutoff because we have not seen more than three different dopant depths in a single image and this depth corresponds roughly to the radius of the features seen at the surface.  We are wary of comparing the strength of any given band bending feature to that of a feature in another image, even when taken with the same tip, since it is subject to frequent minor changes during imaging.  We rule out other non-phosphorus sources of the signature based on the low impurity levels in the wafers and the fact that adsorbed surface contamination, even hydrogen, would show a local topographic effect.  Finally, the appearance of the features as an enhancement in both biases is consistent with what is observed for \textit{n}-type dopants on compound semiconductor surfaces\cite{EbertLong,Zheng94}.  For those materials, the negative sample bias images show an enhancement because the conduction band begins to become occupied as the positive charge bends it near to the Fermi level.  This extra occupied state density in the vicinity of the charge contributes more strongly to the tunneling current, since it is closer to the Fermi level, and produces an enhancement instead of a depression.  Based on all of these considerations we assign these features to buried phosphorus dopants.

The difference between the filled-state signatures of the positively charged dopants and positively charged surface defects may be attributed to the disruption of the surface states by the dimer vacancies.  The buried dopants will obviously allow the antibonding surface band that is derived from the $\pi$ bonded dimer rows to remain intact.  This band is partially resonant with the conduction band but also protrudes into the bulk band gap\cite{Cricenti95}.  As such, at the surface it can act as an extension of the conduction band \cite{McEllistrem93} and provide filled state density at negative sample bias in the same way the conduction band edges do for compound semiconductors \footnote{The surface photovoltage results in reference [12] suggest that the surface band acts as an extension of the conduction band into the gap.  In that work, the photo-induced band bending on Si(100)-(2$\times$1) is not affected by the defect band, which is fully within the gap.  However, the band bending does stop when the antibonding surface states reach the Fermi level, as it does when the band edges read the Fermi level on materials without surface states.}.  However, for dimer vacancies, which break the surface periodicity, the antibonding surface band cannot remain intact, resulting in less protrusion into the band gap, so that an equivalent amount of band bending does not induce filled state density at negative sample bias. 

\section{Conclusion}

In summary, we have identified subsurface dopants near the clean (100)-(2$\times$1) surface of \textit{n}-type silicon using STM.  The charge-induced band bending signature associated with the dopants shows up as an enhancement in both filled and empty states and is consistent with the appearance of \textit{n}-type dopants on compound semiconductor surfaces and passivated Si(100)-(2$\times$1).  This suggests that the antibonding surface states derived from the reconstructed surface dimers act as an extension of the bulk bands into the gap.  Dimer vacancies at the surface appear as depressions in the filled states because they disrupt these surface bands. This result could not be inferred by comparison with previous compound semiconductor work since the vacancy defects observed for those \textit{n}-type materials are always negatively charged.

\begin{acknowledgments}
	This work was supported by the United States Department of Energy under contract number W-7405-ENG-36.  

\end{acknowledgments}

\bibliography{BrownPdopant}% Produces the bibliography via BibTeX.

\begin{thebibliography}{12}
\expandafter\ifx\csname natexlab\endcsname\relax\def\natexlab#1{#1}\fi
\expandafter\ifx\csname bibnamefont\endcsname\relax
  \def\bibnamefont#1{#1}\fi
\expandafter\ifx\csname bibfnamefont\endcsname\relax
  \def\bibfnamefont#1{#1}\fi
\expandafter\ifx\csname citenamefont\endcsname\relax
  \def\citenamefont#1{#1}\fi
\expandafter\ifx\csname url\endcsname\relax
  \def\url#1{\texttt{#1}}\fi
\expandafter\ifx\csname urlprefix\endcsname\relax\def\urlprefix{URL }\fi
\providecommand{\bibinfo}[2]{#2}
\providecommand{\eprint}[2][]{\url{#2}}

\bibitem[{\citenamefont{Tucker and Shen}(2000)}]{TuckerShen00}
\bibinfo{author}{\bibfnamefont{J.}~\bibnamefont{Tucker}} \bibnamefont{and}
  \bibinfo{author}{\bibfnamefont{T.-C.} \bibnamefont{Shen}},
  \bibinfo{journal}{Int.\ J.\ Circ.\ Theor.\ Appl.}
  \textbf{\bibinfo{volume}{28}}, \bibinfo{pages}{553} (\bibinfo{year}{2000}).

\bibitem[{\citenamefont{T.D.~Ladd et~al.}(2002)\citenamefont{T.D.~Ladd,
  Yamaguchi, Yamamoto, Abe, and Itoh}}]{Ladd02}
\bibinfo{author}{\bibfnamefont{J.~G.} \bibnamefont{T.D.~Ladd}},
  \bibinfo{author}{\bibfnamefont{F.}~\bibnamefont{Yamaguchi}},
  \bibinfo{author}{\bibfnamefont{Y.}~\bibnamefont{Yamamoto}},
  \bibinfo{author}{\bibfnamefont{E.}~\bibnamefont{Abe}}, \bibnamefont{and}
  \bibinfo{author}{\bibfnamefont{K.}~\bibnamefont{Itoh}},
  \bibinfo{journal}{Phys.\ Rev.\ Lett.} \textbf{\bibinfo{volume}{89}},
  \bibinfo{pages}{17901} (\bibinfo{year}{2002}).

\bibitem[{\citenamefont{O'Brien et~al.}(2001)\citenamefont{O'Brien, Schofield,
  Simmons, Clark, Dzurak, Curson, Kane, McAlpine, Hawley, and
  Brown}}]{OBrien01}
\bibinfo{author}{\bibfnamefont{J.}~\bibnamefont{O'Brien}},
  \bibinfo{author}{\bibfnamefont{S.}~\bibnamefont{Schofield}},
  \bibinfo{author}{\bibfnamefont{M.}~\bibnamefont{Simmons}},
  \bibinfo{author}{\bibfnamefont{R.}~\bibnamefont{Clark}},
  \bibinfo{author}{\bibfnamefont{A.}~\bibnamefont{Dzurak}},
  \bibinfo{author}{\bibfnamefont{N.}~\bibnamefont{Curson}},
  \bibinfo{author}{\bibfnamefont{B.}~\bibnamefont{Kane}},
  \bibinfo{author}{\bibfnamefont{N.}~\bibnamefont{McAlpine}},
  \bibinfo{author}{\bibfnamefont{M.}~\bibnamefont{Hawley}}, \bibnamefont{and}
  \bibinfo{author}{\bibfnamefont{G.}~\bibnamefont{Brown}},
  \bibinfo{journal}{Phys.\ Rev.\ B} \textbf{\bibinfo{volume}{64}},
  \bibinfo{pages}{161401} (\bibinfo{year}{2001}).

\bibitem[{\citenamefont{Ditchfield et~al.}(2000)\citenamefont{Ditchfield,
  Llera-Rodriguez, and Seebauer}}]{Ditchfield00}
\bibinfo{author}{\bibfnamefont{R.}~\bibnamefont{Ditchfield}},
  \bibinfo{author}{\bibfnamefont{D.}~\bibnamefont{Llera-Rodriguez}},
  \bibnamefont{and} \bibinfo{author}{\bibfnamefont{E.}~\bibnamefont{Seebauer}},
  \bibinfo{journal}{Phys.\ Rev.\ B} \textbf{\bibinfo{volume}{61}},
  \bibinfo{pages}{13710} (\bibinfo{year}{2000}).

\bibitem[{\citenamefont{Brown et~al.}(2002)\citenamefont{Brown, Grube, Hawley,
  Schofield, Curson, Simmons, and Clark}}]{GWBrown02}
\bibinfo{author}{\bibfnamefont{G.}~\bibnamefont{Brown}},
  \bibinfo{author}{\bibfnamefont{H.}~\bibnamefont{Grube}},
  \bibinfo{author}{\bibfnamefont{M.}~\bibnamefont{Hawley}},
  \bibinfo{author}{\bibfnamefont{S.}~\bibnamefont{Schofield}},
  \bibinfo{author}{\bibfnamefont{N.}~\bibnamefont{Curson}},
  \bibinfo{author}{\bibfnamefont{M.}~\bibnamefont{Simmons}}, \bibnamefont{and}
  \bibinfo{author}{\bibfnamefont{R.}~\bibnamefont{Clark}},
  \bibinfo{journal}{J.\ Appl.\ Phys.} \textbf{\bibinfo{volume}{92}},
  \bibinfo{pages}{820} (\bibinfo{year}{2002}).

\bibitem[{\citenamefont{Brown et~al.}(2003)\citenamefont{Brown, Grube, Hawley,
  Schofield, Curson, Simmons, and Clark}}]{GWBrown03}
\bibinfo{author}{\bibfnamefont{G.}~\bibnamefont{Brown}},
  \bibinfo{author}{\bibfnamefont{H.}~\bibnamefont{Grube}},
  \bibinfo{author}{\bibfnamefont{M.}~\bibnamefont{Hawley}},
  \bibinfo{author}{\bibfnamefont{S.}~\bibnamefont{Schofield}},
  \bibinfo{author}{\bibfnamefont{N.}~\bibnamefont{Curson}},
  \bibinfo{author}{\bibfnamefont{M.}~\bibnamefont{Simmons}}, \bibnamefont{and}
  \bibinfo{author}{\bibfnamefont{R.}~\bibnamefont{Clark}},
  \bibinfo{journal}{J.\ Vac.\ Sci.\ Technol.} \textbf{\bibinfo{volume}{21}},
  \bibinfo{pages}{1506} (\bibinfo{year}{2003}).

\bibitem[{\citenamefont{Lequn et~al.}(2001)\citenamefont{Lequn, Yu, and
  Lyding}}]{Lyding01}
\bibinfo{author}{\bibfnamefont{L.}~\bibnamefont{Lequn}},
  \bibinfo{author}{\bibfnamefont{J.}~\bibnamefont{Yu}}, \bibnamefont{and}
  \bibinfo{author}{\bibfnamefont{J.}~\bibnamefont{Lyding}},
  \bibinfo{journal}{Appl.\ Phys.\ Lett.} \textbf{\bibinfo{volume}{78}},
  \bibinfo{pages}{386} (\bibinfo{year}{2001}).

\bibitem[{\citenamefont{Lequn et~al.}(2002)\citenamefont{Lequn, Yu, and
  Lyding}}]{Lyding02}
\bibinfo{author}{\bibfnamefont{L.}~\bibnamefont{Lequn}},
  \bibinfo{author}{\bibfnamefont{J.}~\bibnamefont{Yu}}, \bibnamefont{and}
  \bibinfo{author}{\bibfnamefont{J.}~\bibnamefont{Lyding}},
  \bibinfo{journal}{IEEE Trans.\ on Nanotech.} \textbf{\bibinfo{volume}{1}},
  \bibinfo{pages}{176} (\bibinfo{year}{2002}).

\bibitem[{\citenamefont{Ebert}(1999)}]{EbertLong}
\bibinfo{author}{\bibfnamefont{P.}~\bibnamefont{Ebert}},
  \bibinfo{journal}{Surf.\ Sci.\ Rep.} \textbf{\bibinfo{volume}{33}},
  \bibinfo{pages}{121} (\bibinfo{year}{1999}).

\bibitem[{\citenamefont{Zheng et~al.}(1994)\citenamefont{Zheng, Liu, Newman,
  Weber, Ogletree, and Salmeron}}]{Zheng94}
\bibinfo{author}{\bibfnamefont{J.}~\bibnamefont{Zheng}},
  \bibinfo{author}{\bibfnamefont{X.}~\bibnamefont{Liu}},
  \bibinfo{author}{\bibfnamefont{N.}~\bibnamefont{Newman}},
  \bibinfo{author}{\bibfnamefont{E.}~\bibnamefont{Weber}},
  \bibinfo{author}{\bibfnamefont{D.}~\bibnamefont{Ogletree}}, \bibnamefont{and}
  \bibinfo{author}{\bibfnamefont{M.}~\bibnamefont{Salmeron}},
  \bibinfo{journal}{Phys.\ Rev.\ Lett.} \textbf{\bibinfo{volume}{72}},
  \bibinfo{pages}{1490} (\bibinfo{year}{1994}).

\bibitem[{\citenamefont{Cricenti et~al.}(1995)\citenamefont{Cricenti, Purdie,
  and Reihl}}]{Cricenti95}
\bibinfo{author}{\bibfnamefont{A.}~\bibnamefont{Cricenti}},
  \bibinfo{author}{\bibfnamefont{D.}~\bibnamefont{Purdie}}, \bibnamefont{and}
  \bibinfo{author}{\bibfnamefont{B.}~\bibnamefont{Reihl}},
  \bibinfo{journal}{Surf.\ Sci.} \textbf{\bibinfo{volume}{331-333}},
  \bibinfo{pages}{1033} (\bibinfo{year}{1995}).

\bibitem[{\citenamefont{McEllistrem et~al.}(1993)\citenamefont{McEllistrem,
  Haase, Chen, and Hamers}}]{McEllistrem93}
\bibinfo{author}{\bibfnamefont{M.}~\bibnamefont{McEllistrem}},
  \bibinfo{author}{\bibfnamefont{G.}~\bibnamefont{Haase}},
  \bibinfo{author}{\bibfnamefont{D.}~\bibnamefont{Chen}}, \bibnamefont{and}
  \bibinfo{author}{\bibfnamefont{R.}~\bibnamefont{Hamers}},
  \bibinfo{journal}{Phys.\ Rev.\ Lett.} \textbf{\bibinfo{volume}{70}},
  \bibinfo{pages}{2471} (\bibinfo{year}{1993}).

\end{thebibliography}

\end{document}